\documentclass[aps,prapplied,preprint,amsmath,amssymb,superscriptaddress]{revtex4-1}
\usepackage{graphicx} 
\usepackage[resetlabels,labeled]{multibib}
\newcites{Methods}{Methods}
\usepackage{braket}
\usepackage{hyperref}
\usepackage{amsmath} 
\usepackage{subfig}
\usepackage{siunitx}
\usepackage{booktabs}

\begin{document}
\title{Sympathetic Cooling in Trapped Ions with Spectral Selectivity via the Zeeman
Shift }
\author{Kavyashree Ranawat}
\email{kavyashree.ranawat@duke.edu}
\affiliation{Duke Quantum Center, Duke University, Durham, NC 27701, USA}
\affiliation{Department of Electrical and Computer Engineering, Duke University, Durham, North Carolina 27708, USA}
\author{Jiyong Yu}
\affiliation{Duke Quantum Center, Duke University, Durham, NC 27701, USA}
\affiliation{Department of Electrical and Computer Engineering, Duke University, Durham, North Carolina 27708, USA}
\author{Andrew Van Horn}
\affiliation{Duke Quantum Center, Duke University, Durham, NC 27701, USA}
\affiliation{Department of Electrical and Computer Engineering, Duke University, Durham, North Carolina 27708, USA}
\author{Jacob Whitlow}
\affiliation{IonQ, Inc., College Park, Maryland 20740, USA}
\author{Kenneth R Brown}
\affiliation{Duke Quantum Center, Duke University, Durham, NC 27701, USA}
\affiliation{Department of Electrical and Computer Engineering, Duke University, Durham, North Carolina 27708, USA}
\affiliation{Department of Physics, Duke University, Durham, North Carolina 27708, USA}
\author{Jungsang Kim}
\email{jungsang@duke.edu}
\affiliation{Duke Quantum Center, Duke University, Durham, NC 27701, USA}
\affiliation{Department of Electrical and Computer Engineering, Duke University, Durham, North Carolina 27708, USA}
\affiliation{Department of Physics, Duke University, Durham, North Carolina 27708, USA}

\date{\today}

\vspace{10cm}

\begin{abstract}
    High-fidelity quantum logic operations in trapped ions often require the ions' collective motion to be cooled to near the ground state. Since cooling the ions' motion typically involves dissipative processes such as spontaneous photon scattering, sympathetic cooling is used on select coolant ions between gate sequences to cool the ion chain without affecting the data qubits. Common implementations for coolant ions include different atomic species, different isotopes of the same species or individually addressable ions. Each of these approaches have challenges associated with them, which include increased hardware complexity, reduced efficiency of radial mode cooling and re-ordering events which add additional experimental overhead. We demonstrate a sympathetic cooling scheme leveraging internal metastable atomic levels accessible via a narrow quadrupole transition, utilizing the natural Zeeman shift and individually addressed Raman transitions, to achieve isolation of the non-coolant or ``data ions" from coolant ions. We demonstrate modest decoherence of the data ions due to cooling, while preserving the coherence requirements for high-fidelity gate operations.

\end{abstract}

\maketitle
\section{Introduction}
\label{sec:intr}
Trapped-ions have proven to be one of the most promising technologies for realizing practical quantum computation. With demonstrations of long coherence times \cite{coherence, coherence2}, high fidelity initialization/detection \cite{detect,detect3,detect4} and high fidelity entangling gates \cite{gates,gates2,gates3,gates4},  trapped ions are considered a leading platform for developing fault-tolerant quantum processors \cite{qc1,qc2,qc3,qc4}. There are several challenges associated with scaling system size while maintaining similar levels of controllability and high fidelity logic operations \cite{qc}. Since two-qubit entangling gates often utilize the collective motional modes to mediate interaction between the target ions, motional heating due to noisy electric fields is a main concern in achieving high-fidelity gates \cite{anom_heating,home2009memory, elec_field}. Additionally, some proposals for increasing system size, such as the modular quantum charge-coupled device (QCCD) \cite{qccd1,qccd2} approach involve operations such as chain splitting, shuttling and recombination of ions, which can inject significant motional excitations into the system \cite{shuttling_heat,rowe2002transport}.

Most implementations for high fidelity logic gates \cite{ms} require operation in the Lamb-Dicke regime \cite{wineland1998experimental}, which necessitates motional excitations be kept to minimum levels throughout gate sequences. This is typically achieved by using a set of ``coolant ions" to sympathetically cool the ``data ions", which contain the quantum information, periodically between gate sequences. Due to the Coloumb interaction between the ions in the crystal, the collective modes of motion can be cooled via the coolant ions \cite{larson1986sympathetic}. This results in cooling of the data ions while preserving their internal states, \textit{i.e.} protecting the qubit state. The cooling laser must be decoupled from the internal states of the data ion to ensure that coherence of the ion is minimally affected \cite{be-mg_cool,home2009memory}. For this reason, chains comprising different atomic species are generally used \cite{be-mg_cool, be-al, ca-al}, allowing the use of separate lasers for cooling and gate operations that are very far-detuned from the data ion transitions. Another approach is to use different isotopes and utilize electro-optic modulators to account for the frequency differences between the isotopes that secure some level of isolation \cite{diff_isotope, diff_iso2}. The first method is resource intensive and introduces increased experimental complexity. Owing to the mass differences between species, cooling of radial modes tends to be less efficient from decreased mode participation of the coolant ions \cite{single_ion, ca-al, symp_cool_radial_mode, symp_cool_diff_mass}. This effect is less severe in the second case, however, both approaches suffer from chain re-ordering events upon collision with the background gas molecules, which will require additional shuttling operations to reach the desired chain configuration.

Same-isotope cooling can be employed to circumvent these challenges. The most straightforward approach to realize this is via individual addressing of the cooling ions. However, this can be experimentally challenging to implement, especially in longer chains, whilst ensuring minimal crosstalk on neighboring ions \cite{home2009memory, be-mg_cool}.
In light of these challenges and considerations, we propose and demonstrate a same-isotope cooling scheme, drawing from elements in the 
optical-metastable-ground state qubit (\textit{omg}) architecture \cite{omg}. Spectral isolation of the qubit in this scheme is achieved by using the natural Zeeman shift in our hyperfine qubit along with individual addressing with the Raman beam.

This paper is organized as follows. An outline of the experimental set-up and cooling protocol is provided in Section \ref{sec:setup}. In Section \ref{sec:exp}, selective sympathetic cooling of a target radial mode for a 5-ion chain is demonstrated. Additionally, we target cooling of the radial center-of-mass (COM) mode and tilt mode in the Lamb-Dicke limit, which are more susceptible to heating than others and show that the thermal motion of both modes can be suppressed to near ground state. Importantly, we show evidence of modest impact to the internal coherences of all other non-coolant ions during the application of these cooling pulses. Finally, we propose some measures and modifications to further extend coherence times in Section \ref{sec:outlook}.

\section{Experimental Set-up and Cooling Protocol}
\label{sec:setup}
The experimental details of the cryogenic set-up are provided in Ref \cite{cryo}, with some important hardware upgrades and additions to enable same-isotope cooling. We trap ytterbium ($^{171}$Yb$^{+}$) ions in a surface trap \cite{peregrine} (Peregrine traps from Sandia National Labs), installed in a closed-cycle cryostat (S200, Montana Instruments) which maintains internal temperatures of the trap ``cryopackage" at \SI{8}{\kelvin}. A Q-switched, frequency-doubled Nd:YAG laser at \SI{532}{\nano\meter} is directed at a neutral Yb target for ablation loading and generates a flux of neutral Yb. The two-step photo-ionization of the neutral flux consists of \SI{399}{\nano\meter} laser light resonant with the $^2S_1 \leftrightarrow{} ^2P_1$ transition, followed by \SI{355}{\nano\meter} light to promote the electron to continuum. Essential operations such as initialization, Doppler cooling and state detection are performed using \SI{370}{\nano\meter} light, along with the ``re-pump" laser at \SI{935}{\nano\meter}  to bring the population back into the cycling transition upon decay to the $^2D_{3/2}$ state. An acousto-optic deflector (AOD) based beam steering system \cite{aod1,aod2} allows steering of two tightly focused \SI{355}{\nano\meter} individual beams to selectively address any ion with minimal crosstalk in chains of up to 30 ions \cite{aod}.

The primary cooling laser used in this protocol is a \SI{435}{\nano\meter} laser which addresses the narrow  $^2S_{1/2}
\leftrightarrow {}^2D_{3/2}$ quadrupole transition in $^{171}$Yb$^{+}$ (with a natural linewidth of 3 Hz). It is locked to a stable Fabry–Perot cavity (Stable Laser System, 6010-4), featuring a linewidth of $<$5kHz using the Pound-Drever-Hall (PDH) scheme \cite{pdh1} and is directed at a $45 ^\circ$ angle with respect to the trap axis. This orientation allows it to couple to both the axial and radial modes of motion. The cooling scheme is based on pulsed side-band cooling (SBC), with the laser tuned to the red side-band (RSB) of the $^2S_{1/2}, F=1
\leftrightarrow {}^2D_{3/2}, F=1$ transition. 

The detection scheme employed for determining the transition frequencies on $^2S_{1/2}
\leftrightarrow {}^2D_{3/2}$ line is shown in Figure 1a. This scheme is also used for the cooling demonstration presented in Section \ref{sec:exp}. In the following discussion, we will use the notation $\ket{0}$ and $\ket{1}$ for the qubit states $^2S_{1/2}, F=0$ and $^2S_{1/2}, F=1, m_F = 0$ respectively. All other states will be denoted as $\ket{L,F,m_F}$, where $L$ is the orbital angular momentum, $F$ is the total angular momentum and $m_F$ is its projection along the quantization axis. After the ion is Doppler cooled and initialized, a microwave (MW) $\pi$-pulse is applied between the hyperfine qubit states, bringing the population to the $\ket{1}$ state. Next, it is shelved to the $^2D_{3/2}$ state using the \SI{435}{\nano\meter} laser, with the re-pump laser turned off, and another MW $\pi$-pulse is applied. Finally, the re-pump laser is enabled again and \SI{370}{\nano\meter}-based detection is performed as usual by shining resonant \SI{370}{\nano\meter} light on the $^2S_{1/2}$ to $^2P_{1/2}$ transition. If the \SI{435}{\nano\meter} laser is not resonant with any transition then the second microwave pulse will restore the population to the $\ket{0}$ and the ion will remain dark. Thus, by using this sequence, we can detect the resonance frequencies between the desired states. Figure 1b shows the corresponding spectroscopy results on the $^2S_{1/2} \leftrightarrow {}^2D_{3/2}$ line using this procedure.

\begin{figure}
    \subfloat{
        {\textbf{a}}
        \includegraphics[width = 0.46\linewidth, height = 70mm]{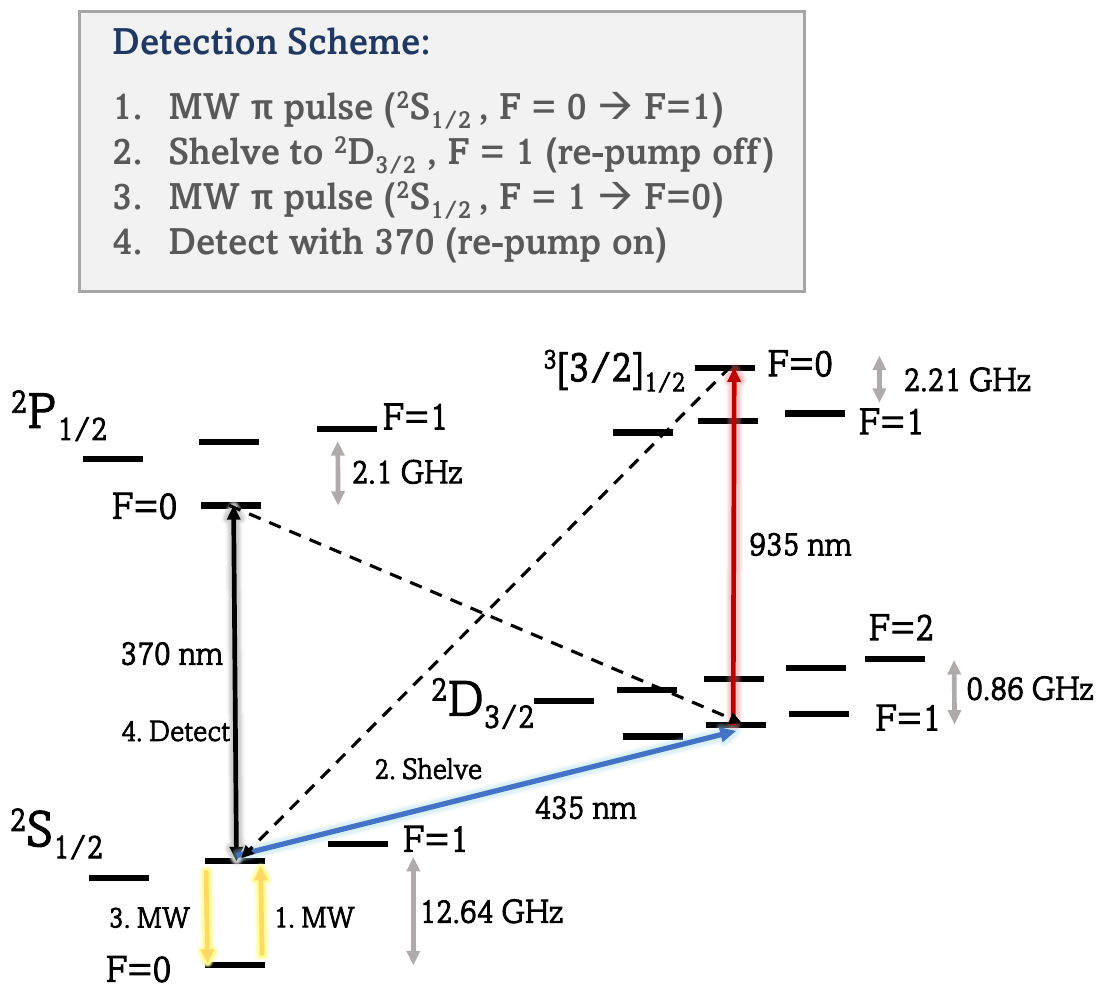} }
        \hspace{1mm}
        \subfloat{
        {\textbf{b}}
        \includegraphics[width = 0.45\linewidth]{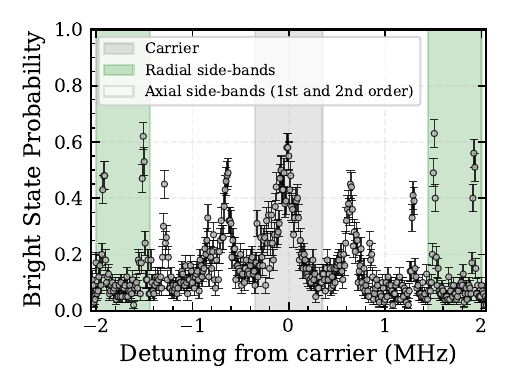}}

    \caption{ \textbf{a}  Detection scheme designed for extracting the resonance frequencies of the ion population shelved to the $^2{}D_{3/2}$ state.\ \textbf{b} Spectroscopy on the $\ket{1} \rightarrow \ket{D,1,0}$ transition, showing the carrier peak (highlighted in grey) in the middle, with the symmetric $1^{st}$ and $2^{nd}$ order axial modes (shown in the white region), and radial modes (highlighted in green).}
\end{figure}

\begin{figure*}
        \subfloat{{\textbf{a}}{\includegraphics[height = 60mm,width=.46\linewidth]{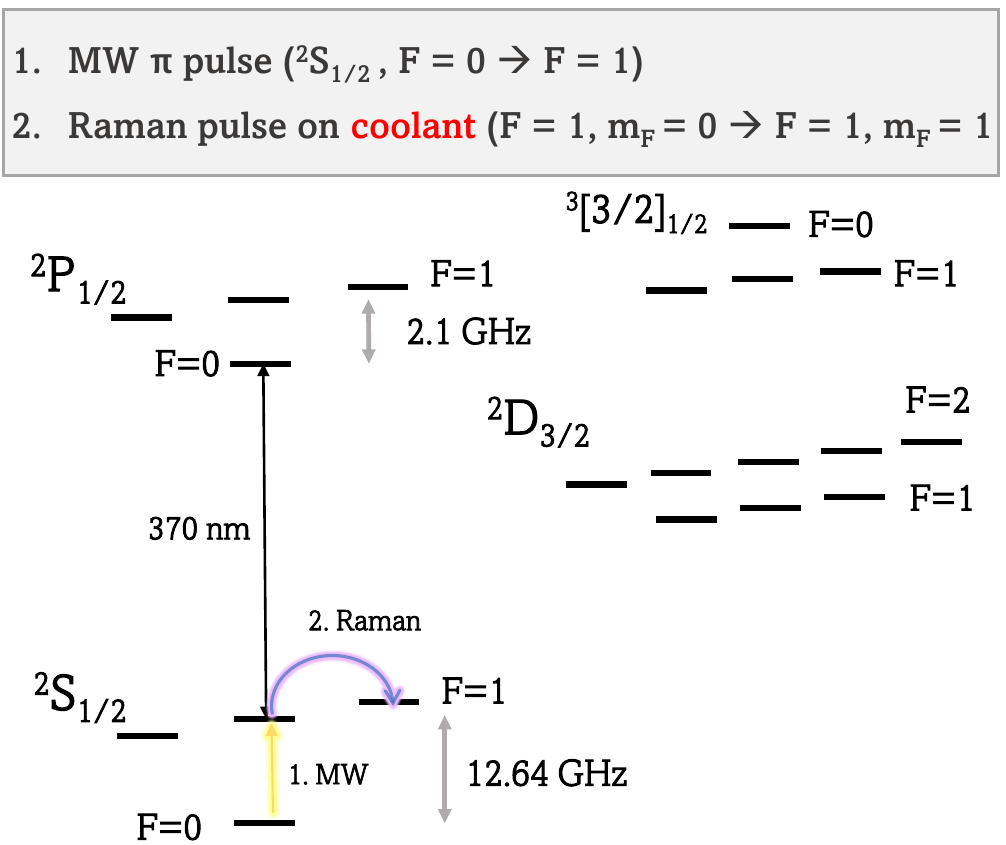}}%
            \label{subfig:a}%
        }\hfill 
        \subfloat{{\textbf{b}}{%
            \includegraphics[height = 60mm,width=.45\linewidth]{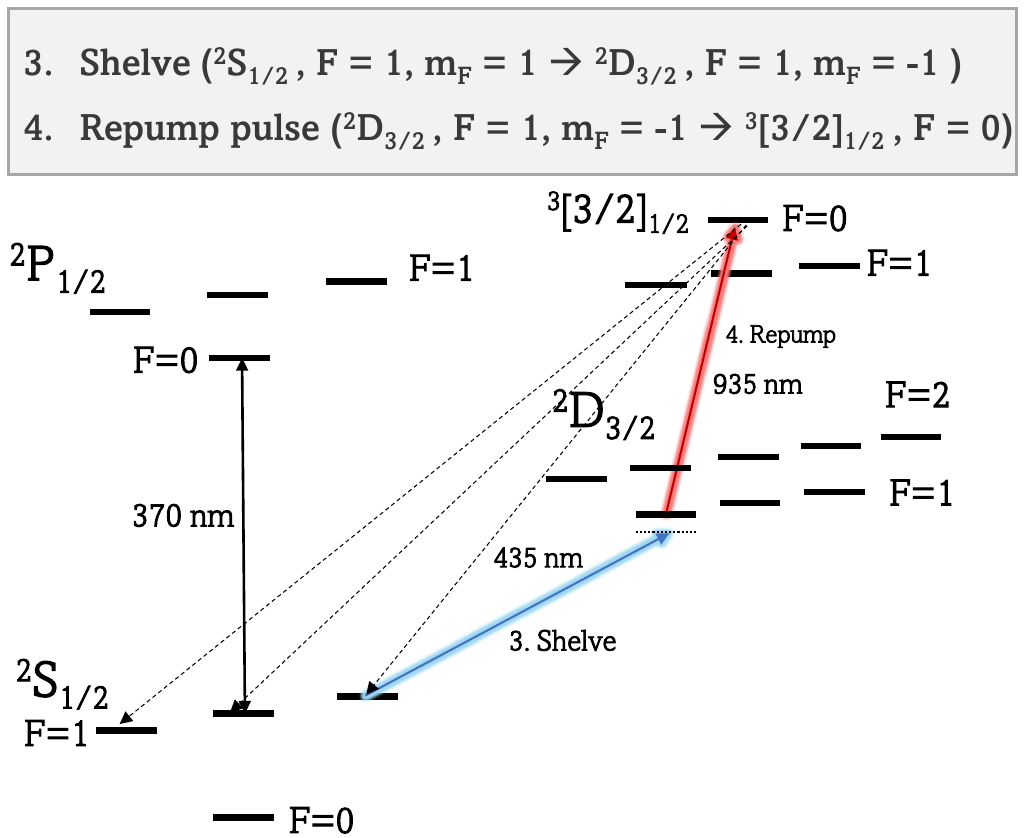}%
            \label{subfig:b}%
        }}\\
        \subfloat{{\textbf{c}}{%
            \includegraphics[height=60mm,width=.45\linewidth]{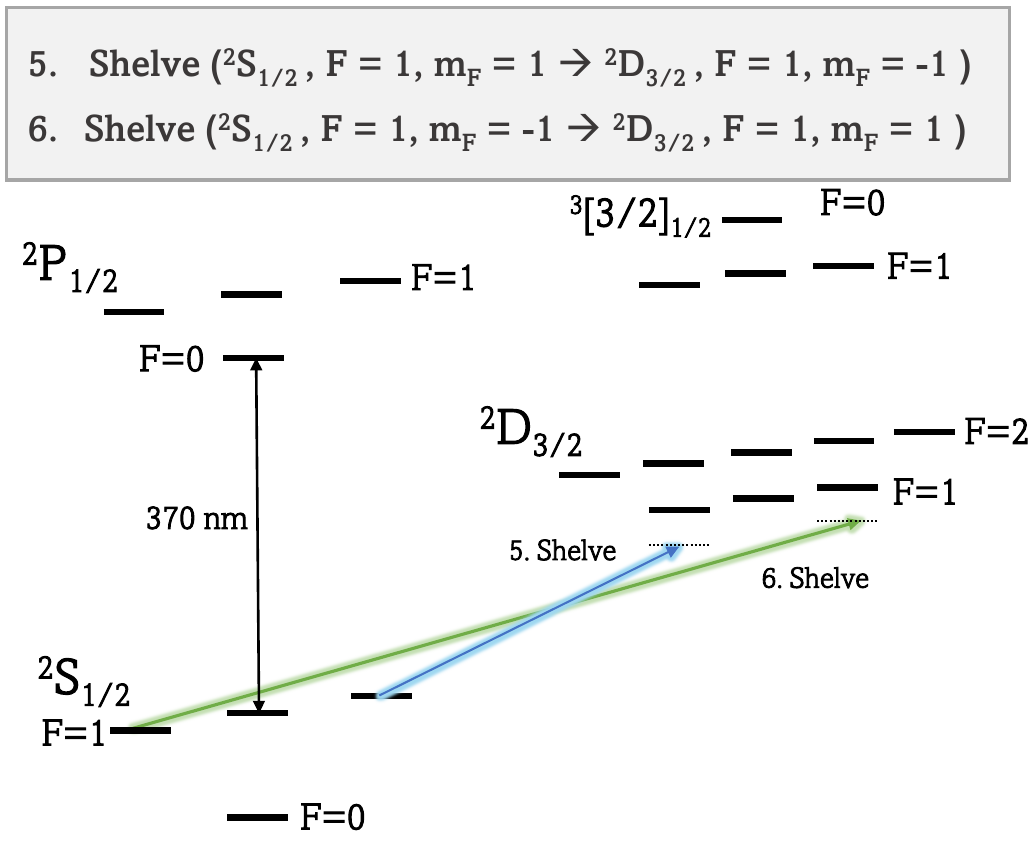}%
            \label{subfig:c}%
        }}\hfill
        \subfloat{{\textbf{d}}{%
            \includegraphics[height=60mm,width=.45\linewidth]{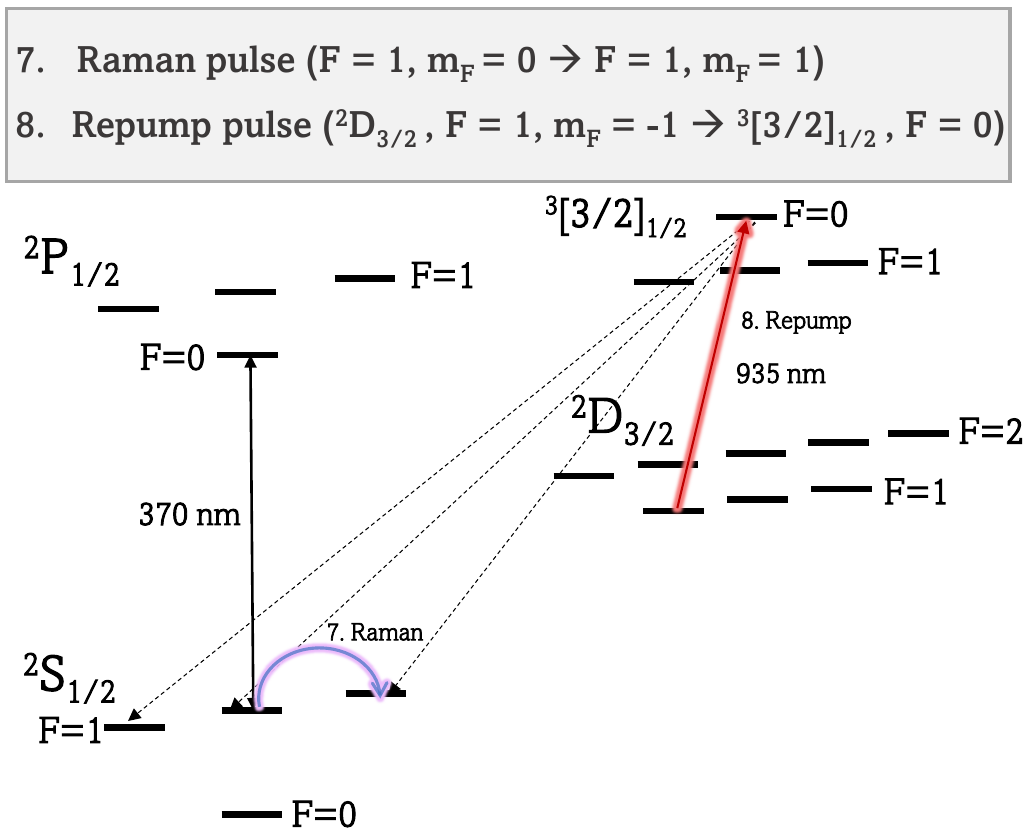}%
            \label{subfig:d}%
        }}
        \caption{Same-isotope sympathetic cooling sequence for  $^{171}$Yb$^{+}$. \textbf{a} Ions are initialized in the ground state of the S-manifold, following which a MW-$\pi$ pulse is applied between the hyperfine states. The individual Raman beam selectively addresses the coolant ion to bring it to the upper Zeeman state by applying two RF tones at a beat frequency equal to the Zeeman shift. A second MW-$\pi$ pulse brings the data ion back to the ground state. \ \textbf{b} After initializing the coolant, it is shelved to the $\ket{D,1,-1}$ state at the RSB frequency. The re-pump laser at \SI{935}{\nano\meter} pumps the population to the ${}^3[3/2]_{3/2}, F=0$ bracket state, followed by spontaneous emission back to all the states in $\ket{S,1}$ level.  \ \textbf{c} The population in the upper Zeeman state is again shelved at the RSB frequency to $\ket{D,1,-1}$. Subsequently, population in the lower Zeeman state is shelved at the RSB frequency to $\ket{D,1,+1}$. \ \textbf{d} The remainder of the population is now entirely in the $\ket{1}$ hyperfine state, which is transferred again to the upper Zeeman state via the individual Raman beam. The re-pump laser is then enabled again, which results in the population re-setting to the $\ket{S,1}$ levels. This completes 1 cooling cycle.}
        
    \end{figure*}

The same-isotope cooling sequence is presented in Figure 2. An important feature in this scheme is the selective addressing of the coolant ion with the individually-addressed Raman beam at \SI{355}{\nano\meter} using the AOD system. Consider the basic case of only two ions, where one is the coolant and the other the data ion. First, a MW $\pi$-pulse is performed between the qubit states, leaving both ions in the $\ket{1}$ state. Next, two co-propagating RF tones are applied onto the Raman individual beam, with the frequency difference between the tones tuned to drive  the $\ket{1}\leftrightarrow \ket{S,1,+1}$ transition for the coolant ion (Figure 2a). This pulse moves the electron in the coolant to the upper Zeeman state $\ket{S,1,+1}$, separated by a Zeeman shift of \SI{11}{\mega\hertz} in our set-up. Another MW $\pi$-pulse is then performed to return the data ion to its qubit ground state. For the cooling cycle, the coolant ion is first shelved from the Zeeman state with the \SI{435}{\nano\meter} laser frequency tuned to first RSB of $\ket{D,1,-1}$, followed by enabling the re-pump laser between $\ket{D,1}\leftrightarrow {}^3[3/2]_{3/2}, F=0$ (Figure 2b). At this stage, the population decays to all the sub-levels in the $\ket{S,1}$ manifold. It is important to note that population decay to $\ket{0}$ is forbidden, which eliminates the need for any additional pulses to remove cooling population from this state. The population in both the Zeeman levels is then shelved sequentially using the \SI{435}{\nano\meter} laser, by driving a RSB $\pi$-pulse on the $\ket{S,1,+1} \rightarrow \ket{D,1,-1}$ transition, followed by a RSB $\pi$-pulse on the $\ket{S,1,-1} \rightarrow \ket{D,1,+1}$ transition (Figure 2c). With this, only population in $\ket{1}$ remains, which is again transferred to the upper Zeeman state via the individually-addressed Raman pulse described earlier. Finally, the re-pump laser is enabled, causing decay to all the sub-levels in the $\ket{S,1}$ manifold once again (Figure 2d). The combination of the two RSB shelving pulses, the Raman individual pulse and the re-pump pulse results in cooling and comprises one cooling pulse sequence. Repeating this pulse sequence for sufficient number of cycles should result in ground state cooling of the coolant ion and thereby the data ion through sympathetic cooling. The motivation for shelving from $\ket{S,1,\pm 1} \rightarrow \ket{D,1,\mp 1}$ is so that any carrier transition from the qubit state $\ket{1}$ is detuned by at least the Zeeman shift to reduce the likelihood of off-resonant couplings.

\section{Experimental Results}
\label{sec:exp}

\subsection{Cooling Protocol Implementation}
\label{subsec:rsb_cool}

As a proof-of-principle demonstration, we implement the cooling protocol on the radial modes in a 5-ion chain of $^{171}$Yb$^{+}$ from the Doppler limit. It is important to note, however, that this sympathetic cooling scheme can be deployed between gate sequences when running complete quantum circuits. In this scenario, it is appropriate to operate in the Lamb-Dicke regime, meaning fewer cooling pulses will be required to keep the average Fock state population low. This will be discussed further in Section \ref{subsec:heating_rate}, where cooling is deployed to counteract the heating rate of target modes.

\begin{figure}
\includegraphics[width = 0.7\textwidth, height = 14 cm]{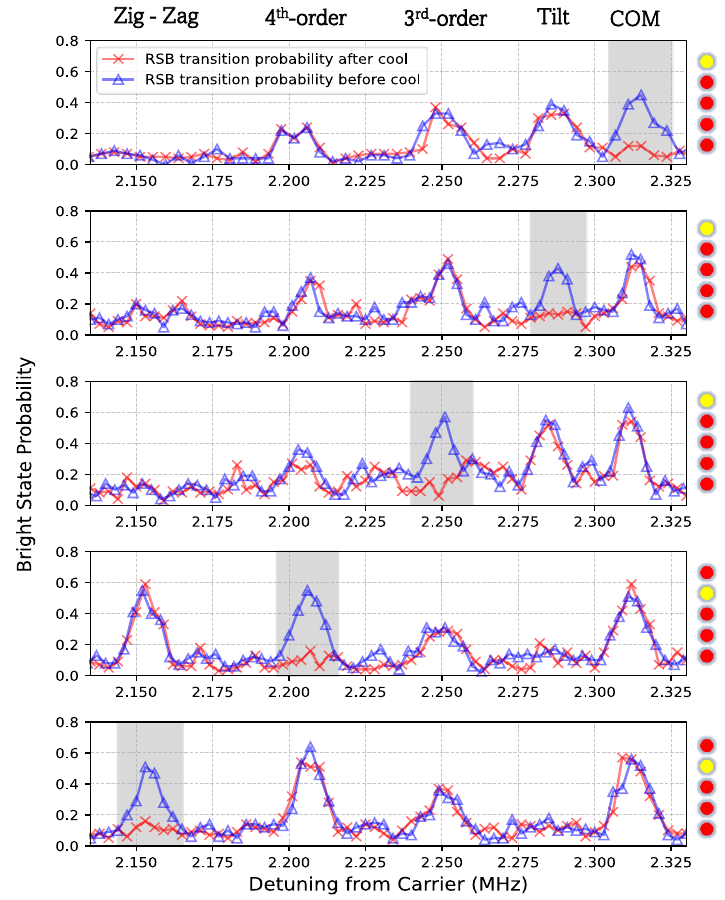}

    \caption{ Demonstration of targeted cooling for all modes in a 5-ion chain of $^{171}$Yb$^{+}$ ions. Following the cooling sequence, an initialization pulse brings the ion population to the ground state. The population is then shelved to the $\ket{D,1,0}$ state at the Doppler-cooled $\pi$-times in order to probe all RSB transition probabilities, with the rightmost peak representing the COM mode. The blue markers here depict the RSB transition probabilities post Doppler cooling while the red markers show those following Doppler cooling and sympathetic cooling, with the mode that is subject to cooling highlighted in grey. The edge ion is designated as the coolant for cooling the COM, tilt and 
    $3^{rd}$ order modes, while the ion adjacent to the center ion is used for cooling the $4^{th}$ order and zig-zag modes. Each mode is labeled vertically above the corresponding peak and the ion assigned as the probe is depicted in yellow in the 5-ion chain shown alongside each sub-plot. The number of cooling pulses is fixed at 100 for all modes for the purposes of a basic demonstration of the scheme in practice.}
    
\end{figure}

Any ion can be designated as the coolant and used for selectively targeting the modes in which it participates for cooling. After sympathetic cooling with a specific ion, an initialization pulse is performed with the \SI{370}{\nano\meter} as is typically done, bringing all the ions to the ground state. The cooled mode is probed using the shelving detection scheme described in the previous section, specifically, the RSB transition of the $\ket{1} \rightarrow \ket{D,1,0}$ is used. Since the \SI{435}{\nano\meter}  laser is a global beam, with a waist radius of \SI{21}{\micro\meter}, it will excite all the ions simultaneously. In this experiment, we demonstrate sequential cooling of the targeted mode in the higher frequency radial axis, beginning from the highest frequency center-of-mass (COM) mode to the lowest frequency zig-zag mode, as shown in Figure 3. The edge ion is assigned as the coolant ion for cooling the COM, tilt and third order modes. For the fourth order and zig-zag mode, the ion adjacent to the center assumes the role of coolant by steering the AOD frequency to address it. The number of cooling pulses is fixed at 100 for all modes, with the pulse time varying between $270-295\ \mu s$ per pulse depending on the calibrated $\pi$ times of the shelving and Raman pulses. The blue markers indicate the RSB transition probabilities after Doppler cooling but prior to sympathetic cooling, whereas the red markers indicate the RSB transition probabilities measured after applying the same-isotope cooling protocol. The grey shaded region represents the target mode in each experiment, with corresponding RSB mode suppression following cooling indicating motional modes in near-ground state. The average motional population, $\bar{n}$ measured via the side-band thermometry method \cite{ratio} is measured to be less than 1 for all the modes.

\subsection{Heating Rate Suppression}
\label{subsec:heating_rate}
To perform a series of gates with high fidelity, periodic application of cooling pulses can be used to suppress the heating rate of modes susceptible to heating. In the 5-ion chain, the baseline heating rate of the COM mode is 30 quanta per second (q/s), the highest of all other modes. This is the only mode that is readily excited by uniform electric field noise, while the other modes can only be excited by higher-order field gradients, leading to much lower heating rates~\cite{com_cool2}.  Although we often perform logical gates using the ``cooler modes", heating on the COM mode can indirectly affect the logical mode by causing unwanted fluctuations in the Rabi frequency, thus limiting gate fidelity \cite{com_cool,com_cool2}. For this reason, we primarily target cooling of the COM mode. We also show cooling performance for the tilt mode, which has the second highest heating rate of $3-4$ q/s on average.

\begin{figure}
    \subfloat{
        {\textbf{a}}
        \includegraphics[width = 0.46\linewidth, height=55mm]{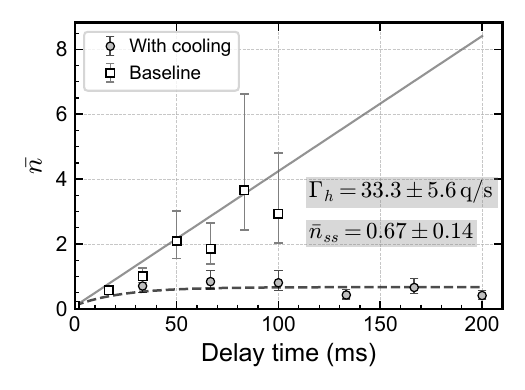}}
        \subfloat{
        {\textbf{b}}
        \includegraphics[width = 0.46\linewidth, height = 55mm]{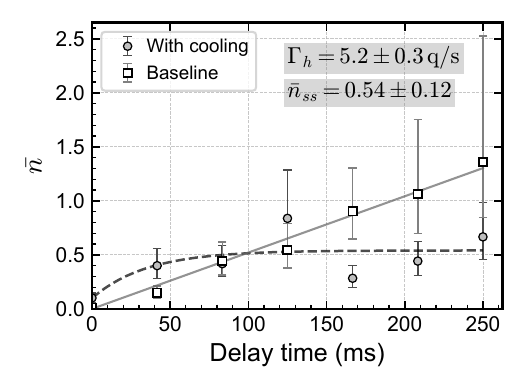}}
    \caption{ Heating rate suppression of the COM and tilt modes in a 5-ion $^{171}$Yb$^{+}$ chain. The square markers represent the average motional quanta for the baseline case while the circular markers show the average motional population after periodic cooling for both modes. Each cooling cycle consists of two shelving transitions between Zeeman states, one Raman pulse and one re-pump pulse.  \textbf{a} To cool the COM mode, 10 cooling pulses are injected, followed by an idling time such that the total period is \SI{10}{\milli\second}. This leads to an equilibrium point where a steady state limit of $\bar{n}$ is reached, which in this case is $0.67 \pm 0.14$ q. \textbf{b} For cooling the tilt mode, cooling is performed at a periodicity of 10 pulses within a \SI{50}{\milli\second} time period, leading to a steady state limit of $\bar{n} = 0.54 \pm 0.12$ q in this demonstration. The baseline heating rate for the COM and tilt mode extracted from linear fits shown above is $\Gamma_h = 33.3 \pm 5.6$ q/s and $\Gamma_h = 5.2 \pm 0.12$ q/s respectively.}
\end{figure}

For this demonstration, 10 cooling pulses for the COM mode are injected in a \SI{10}{\milli\second} time period. The choice for this periodicity was informed by evaluating the cooling performance of different combinations of pulses and delay times as well as simulations. Figure 4a shows $\bar{n}$, illustrated with square markers for the baseline case and  circular markers for $\bar{n}$ for the case where periodic cooling is applied. For the cooling case, 10 cooling pulses are applied, followed by an idling time such that the total period of cooling and idling cycle is \SI{10}{\milli\second}. The initial pulses result in a decrease in heating rate, until eventually an equilibrium is reached where the cooling rate matches the heating rate at a given $\bar{n}$. In this example, a steady state value of $\bar{n} = 0.67 + 0.14$ q is reached after $\leq$ \SI{200}{\milli\second}. We note that the baseline heating rate affects the overall achievable steady state $\bar{n}$. For example, depending on the baseline heating rate, the steady state $\bar{n}$ can vary anywhere between 0.3-0.7 q. As mentioned previously, each cycle comprises the two RSB shelving transitions, a copropagating Raman pulse and the re-pump pulse to pump the cooled population back to the $\ket{S,1}$ states. Each pulse is applied for \SI{75}{\micro\second}, \SI{135}{\micro\second}, \SI{65}{\micro\second}  and \SI{20}{\micro\second} respectively, resulting in a total cooling duration of \SI{2.95}{\milli\second} for 10 pulses. Figure 4b demonstrates heating rate suppression for the tilt mode, which can sustain a longer wait time before cooling pulses need to be injected. In this case, 10 pulses are used in a \SI{50}{\milli\second} cooling period to achieve a steady state $\bar{n}$ of $\approx$ 0.5 q. The total cooling time for 10 pulses for the tilt mode is \SI{3.25}{\milli\second}.

The effect of the duty cycle on the cooling performance is also investigated. Here, we define the duty cycle as the idling time during which actual gate operations can take place with respect to the total time that includes cooling. For example, in the above demonstration, the  duty cycle used to suppress the heating rate to an $\bar{n} < 1$ for the COM mode is $70.5 \% (\frac{\SI{10}{\milli\second} \text{ - } \SI{2.95}{\milli\second}}{\SI{10}{\milli\second}})$, whereas for the tilt mode is $93.5 \%  (\frac{\SI{50}{\milli\second} \text{ - }  \SI{3.25}{\milli\second}}{\SI{50}{\milli\second}})$. Although this is a useful measure to express the proportion of time assigned for cooling, it should be emphasized that the same cooling performance can be achieved for a higher duty cycle by decreasing the cooling pulse times. For example, using a cooling beam with a smaller beam waist or higher power and a Raman beam with more suitable polarizations for driving to the Zeeman levels will result in faster cooling times and thus a higher duty cycle. Figure 5 shows the steady state $\bar{n}$ as a function of varying number of cooling pulses and the corresponding duty cycle. For each experiment with a given number of cooling pulses, the equilibrium $\bar{n}$ value is calculated from the average value obtained from five independent experimental runs, represented by the grey markers.
Unsurprisingly, the overall steady state $\bar{n}$ for the COM mode reduces with a lower duty cycle but reaches a steady value at $\approx  12$ pulses, indicating the redundancy of adding more cooling pulses. Cooling on the tilt mode exhibits a similar trend with marginal improvements in the steady state limit as the number of cooling pulses is increased from 5 to 30. From these results, we can infer cooling strategies depending on the system requirements. For example, if the aim is to keep both the COM and tilt modes below a target threshold, say $\bar{n}<1$, then the maximum achievable duty cycle to maintain this condition is $<87\%$.
\begin{figure}
\subfloat{
        {\textbf{a}}
        \includegraphics[width = 70mm]{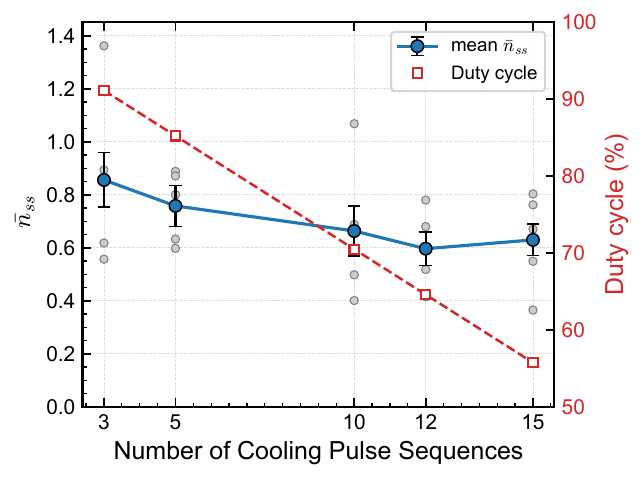}}
        \subfloat{
        {\textbf{b}}
        \includegraphics[width = 70mm]{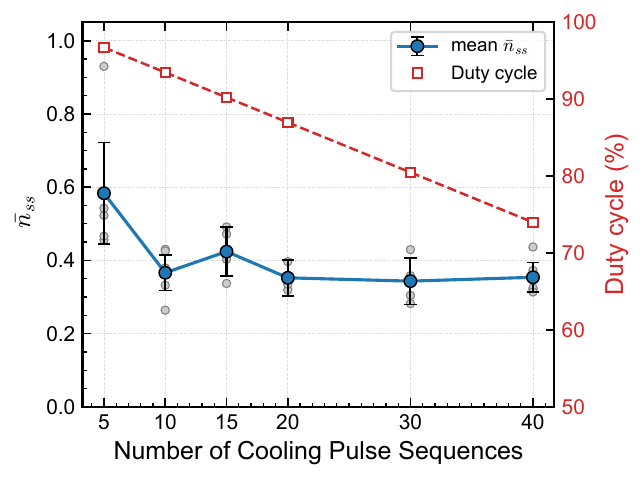}}

    \caption{ Steady state average phonon number versus the number of cooling pulses and duty cycle for the COM (\textbf{a}) and tilt (\textbf{b}) modes in a 5-ion chain of $^{171}$Yb$^{+}$ ions. The duty cycle represents the proportion of idling time during which quantum gate operations can be performed over the total period that also includes cooling time. For each duty cycle, the steady state limit (blue markers) is extracted from an average of 5 experimental runs (grey markers) to get a better representation of the qualitative trend.}
\end{figure}

\subsection{Effect on Coherence of Data Qubits}
\label{subsec:coherence}
In order for the cooling protocol to be practically applicable, the effect of the cooling laser on the coherence of non-coolant ions must be minimized. To probe this effect, Ramsey interferometry driven by MW gates, with a spin-echo sequence for canceling out any static light shifts, was used to measure coherence of the neighboring data qubits while cooling pulses were being actively applied. We apply cooling at a rate of 10 pulses within a \SI{10}{\milli\second} time segment during the wait period of the Ramsey interference, the same conditions utilized to suppress the heating rate for the COM mode. To perform the echo, the second MW $\pi$ pulse used in the cooling protocol (Figure 2a) is replaced with a $\pi/2$ pulse, placing all the non-coolant ions in a superposition state. The sequence of cooling and delay is repeated for the entire duration of the both $\tau/2$ wait times, which surround the echo pulse in the middle. Finally, another MW $\pi/2$ pulse, brings all the non-coolant ions back to the ground state. The baseline coherence measured (without the cooling pulses) is 
$> \SI{2}{\second}$, enabling us to attribute a reduction in coherence time to the application of cooling during the wait period. The results from this coherence measurement are shown in Figure 6. The average $1/e$ time for all the non coolant ions is $\approx$ \SI{650}{\milli\second}, setting a lower bound on the execution time of high-fidelity gate sequences. Although reduced somewhat, the qubit coherence is maintained quite well, with a coherence loss of $2.37 \times 10^{-3}$/s per cooling pulse. Taking into account the duty cycle, this corresponds to an error of $3.5\times10^{-5}$ per \SI{10}{\milli\second} cooling cycle with respect to evolution under the identity gate.  The run times and resulting errors can be further improved by selecting the appropriate cooling strategies and duty-cycles.
\begin{figure}
        \includegraphics[width = 0.5\linewidth,height = 65 mm]{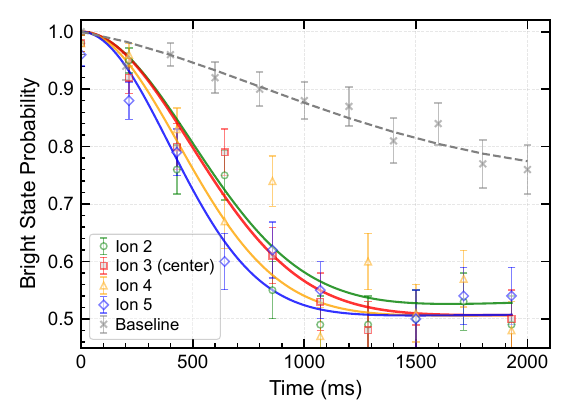} 
    \caption{ Microwave Spin-Echo showing the coherence times for the non-coolant ions while applying cooling at a rate of 10 pulses in \SI{10}{\milli\second} during both $\tau/2$ waiting times. The average coherence times for the non-coolant ions is extracted from fitting the data to an overall Gaussian fit which also includes a slow increase of the steady state population from \SI{370}{\nano\meter} light leakage at much longer wait times. The extracted coherence times are $\SI{712 \pm 83}{\milli\second}$ (ion 2), $\SI{696 \pm 71}{\milli\second}$ (ion 3), $\SI{620 \pm 44}{\milli\second}$ (ion 4) and $\SI{559 \pm 70}{\milli\second}$ (ion 5). This results in an average  $1/e$ time of $\approx$ \SI{650}{\milli\second} across all ions. The baseline MW coherence, depicted by the grey x-markers, is  $\SI{2092 \pm 123}{\milli\second}$.}
\end{figure}

\begin{figure}
\subfloat{
        {\textbf{a}}
        \includegraphics[width = 70mm]{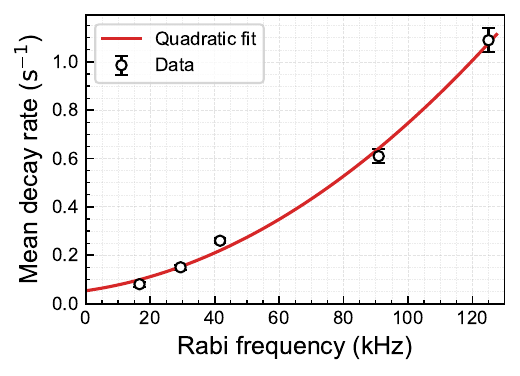}}
        \subfloat{{\textbf{b}}
        \includegraphics[width = 70mm]{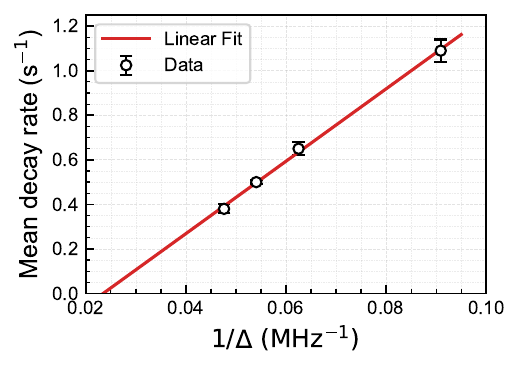}}
    \caption{  \textbf{a}  The average decay rate of all non-coolant ions as a function of \SI{435}{nm} laser power in terms of the Rabi frequency. The data fits to a quadratic curve of the form $ax^2+bx+c$, where $a = 5.03 \times10^{-11}, b = 1.91\times10^{-6}$ and $c=0.55$.\textbf{ b}  The average decay rate of all non-coolant ions as a function of inverse \SI{435}{nm} laser detuning. The data fits to a linear function of the form $mx+c$, where $m = 16.24, c = 0.12 $. In both plots, the error bars for the coherence times are calculated from the standard deviation of a gaussian decay fit. This error is subsequently converted into a decay error (shown here) by Taylor-expanding about the decay rate. The mean decay rate shown here is adjusted so as to remove the contribution of decoherence without the \SI{435}{nm} beam. }
    \label{fig7}
\end{figure}

\begin{figure}
    \includegraphics[width = 0.7\linewidth]{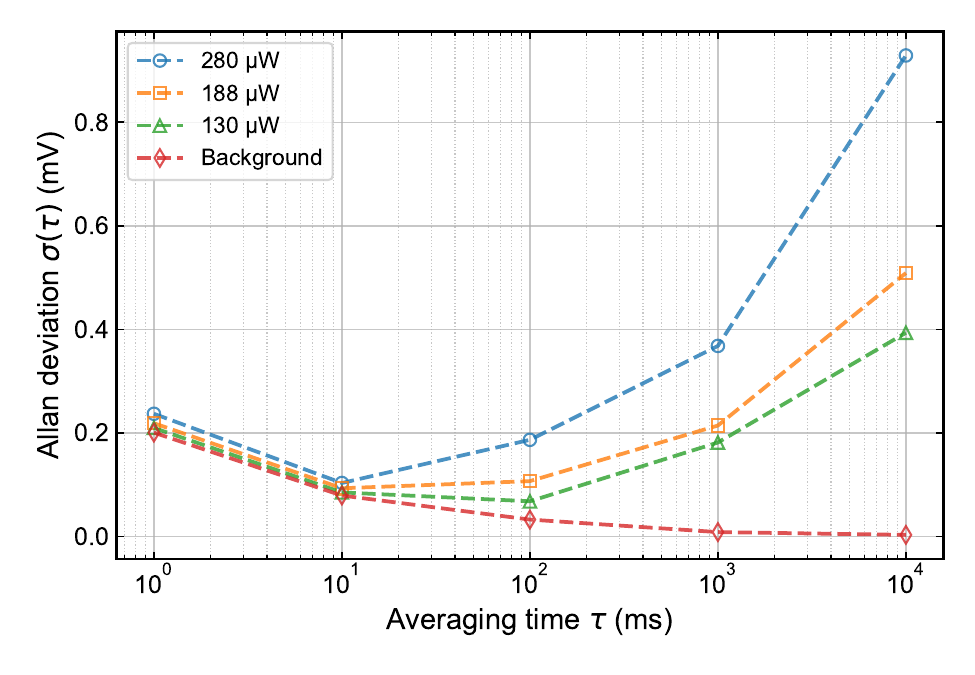} 
    \caption{Allan deviation of the background noise and the \SI{435}{nm} beam amplitude for a total time duration of \SI{100}{\second} for different settings of optical laser power. The deviation measurements shown here are not normalized with respect to the overall mean amplitude in order to show the behavior of the laser amplitude fluctuations with respect to the background signal.}
    \label{fig8}
\end{figure}

The primary source of decoherence originates from the \SI{435}{\nano\meter} beam, which arises from noise in the light shift induced from applying the beam during cooling as its intensity fluctuates. Although the echo sequence cancels out the static light shifts, it cannot compensate for the fluctuations in the light shift that may arise from intensity fluctuations in the laser. To further confirm this, the dependence of the coherence decay rate as a function of both the field strength $\Omega_{435}$ and the detuning $\Delta_{435}$ from the cooling transition was measured. The average decay rate for all non-coolant ions exhibits a quadratic dependence to $\Omega_{435}$ (Fig.~\ref{fig7}a) and inversely proportional to $\Delta_{435}$ (Fig.~\ref{fig7}b), which is consistent with having a time dependent drift in $\Omega_{435}$. We measured the long-term drift of the intensity for the \SI{435}{\nano\meter} laser driving the cooling transition (Fig.~\ref{fig8}), and matched the rate of decoherence reduction to the fluctuation in qubit frequency induced by the light shift. Implementing an intensity lock for the laser to stabilize the fluctuations can further minimize the impact on coherence [See Appendix].

Another potential source of decoherence is the effect of absorption of the \SI{297}{\nano\meter} photons by the non-coolant ions, which are spontaneously emitted on the $\ket{[3/2],0} \rightarrow \ket{S,1}$ transition. This effect was probed experimentally by repeating the coherence measurement with the regular cooling sequence but disabling the \SI{935}{\nano\meter} re-pump laser to eliminate the \SI{297}{\nano\meter} transition. We detected no observable difference in the measured decay rate, and conclude that this is not the dominant source of decoherence as is consistent with the approximate scatter rate calculation shown in the Appendix. 

\section{Outlook}
\label{sec:outlook}
In summary, we have demonstrated a scheme which enables sympathetic cooling in same-isotope chains by utilizing the natural Zeeman shift and a beam that allows for individual addressing of ions in the chain with low impact on the data ions. One extension of this scheme is real-time chain re-configuration of the ions without shuttling. Although a specific coolant is assigned to cool a target mode at the start of the experiment, another mode can be targeted for cooling mid-experiment by swapping the roles of the data ion and coolant.  To implement this, M$\o$lmer-S$\o$rensen gates \cite{Sorensen1999, PhysRevA.62.022311, PhysRevLett.82.1835} can be performed to swap information such that the original data ion assumes the role of the coolant ion. This proposal allows flexibility in designating coolant ions mid-experiment while circumventing the need for physically moving ions to achieve the same.

Further improvements in coherence are possible by incorporating a few measures, including intensity stabilization of the cooling laser and increasing the Zeeman shift to minimize the light shift by increasing the detuning from the cooling transition. As previously mentioned, individual addressing of ions by the cooling beam directly would also eliminate light shift as the beam is shaped such that it is focused only on the coolant ions. 

Another application of the same-isotope cooling scheme for cooling radial modes is in the simulation of open quantum systems. This protocol can be readily used as a source of tunable dissipation \cite{dissipn,disspn2} to simulate models with system-bath couplings while simplifying the hardware constraints that are used to typically achieve this using a second atomic species or different isotope.

\section{Acknowledgments}
This work was supported by the Office of the Director of
National Intelligence - Intelligence Advanced Research Projects Activity through
ARO contract W911NF16-1-0082 (baseline hardware), the National Science Foun-
dation STAQ project (PHY 2325080, sympathetic cooling capability) and the U.S.
Department of Energy Quantum Systems Accelerator (DE-FOA-0002253, protocol
development and experimentation).

\section{Data Availability}
The raw data used to produce the experimental results in the main text can be found here: \url{https://github.com/kr9803k/Same-Isotope-Cooling}.

\section{references}
\bibliography{refs.bib}  
\label{LastBibItem}
\section{Appendix}
\label{sec:appendix}
\subsection {Effect of Axial Coupling on Cooling Performance}
\label{subsec:Axial_Coupling}

One of the factors that affect cooling performance is strong coupling to the axial modes of motion. Due to our \SI{435}{\nano\meter} beam geometry, its projection along the axial direction is $1.4$ times higher than the radial axes. Additionally, the axial COM frequency is \SI{330}{\kilo\hertz}, resulting in a higher Lamb-Dicke factor, $\eta_{ax}$ for the axial modes. Due to the high degree of coupling to these ``hot" modes, the cross terms between the radial and axial modes cannot be neglected. The Hamiltonian considering a single radial mode and axial mode in the interaction picture is given by \cite{single_ion,mutlimode_hamiltonian}

\begin{equation}\label{eq3}
    H = \frac{\hbar\Omega}{2}\hat{\sigma}_+ \text{exp}(-i\Delta t)\text{ exp}\big(i\eta_{ax}(\hat{a}e^{-i\nu_{ax}t}+\hat{a}^{\dagger}e^{i\nu_{ax}t}) + i\eta_{rad}(\hat{b}e^{-i\nu_{rad}t}+\hat{b}^\dagger e^{i\nu_{rad}t})\big) + \text{h.c.} 
\end{equation}

where $\Omega$ is the Rabi frequency of the transition, $\hat{\sigma}_+$ is the raising operator for the atomic transition and $\Delta$ is the laser detuning from the transition. The motional operators and motional frequency for the axial and radial modes are given by $\hat{a} (\hat{a}^{\dagger}) \text{, } \hat{b}(\hat{b}^{\dagger})$ and $\nu_{ax}, \nu_{rad}$ respectively.
Here, we assume the Lamb-Dicke approximation may not hold for the axial mode since the condition $\eta_{ax}^2(2\bar{n}_{ax}+1) \ll 1$ is not strictly satisfied, where $\bar{n}_{ax}$ is the average phonon number for the axial mode. 

Expanding the exponentials up to second order using the Taylor series, we get
\begin{equation}
    \begin{aligned}
        H = \frac{\hbar\Omega}{2}\hat{\sigma}_+ &e^{-i\Delta t}
        \big(1+i\eta_{ax}(\hat{a}e^{-i\nu_{ax}t }+\hat{a}^{\dagger}e^{i\nu_{ax}t}) -\frac{\eta_{ax}^2 }{2}(\hat{a}^2e^{-2i\nu_{ax}t} + \hat{a}\hat{a}^\dagger + \hat{a}^\dagger \hat{a}+ \hat{a}^2e^{2i\nu_{ax}t})\big) \\
        \ & 
        \big( 1+i\eta_{rad}(\hat{b}e^{-i\nu_{rad}t }+\hat{b}^{\dagger}e^{i\nu_{rad}t}) -\frac{\eta_{rad}^2}{2} (\hat{b}^2e^{-2i\nu_{rad}t} + \hat{b}\hat{b}^\dagger + \hat{b}^\dagger\hat{b} + \hat{b}^2e^{2i\nu_{rad}t})\big) + \text{h.c.} 
    \end{aligned}
\end{equation}

Letting the detuning $\delta$ be resonant with the first-order radial RSB and ignoring fast-rotating terms yields the following radial RSB Hamiltonian.

\begin{align}
    H_{RSB,rad} &= \frac{\hbar\Omega}{2}\hat{\sigma}_{+}e^{i\nu_{rad} t}\big(1-\frac{\eta_{ax}^2}{2} ( \hat{a}\hat{a}^\dagger + \hat{a}^\dagger\hat{a})\big)
    \big(i\eta_{rad}\hat{b}e^{-i\nu_{rad}t}\big) + \text{h.c.} \\
    &= \frac{\hbar\Omega}{2}\hat{\sigma}_{+}\big( i\eta_{rad}\hat{b} - i\eta_{rad}\eta_{ax}^2 (\bar{n}_{ax}+\frac{1}{2})\big)
\end{align}

This results in an effective distribution of RSB Rabi frequency as it samples from a distribution which originates from the axial coupling. This could reduce cooling efficiency since the RSB shelving pulses, which are driven at fixed times, will sample different Rabi frequencies for the duration of cooling. 

\subsection{Decay Rate Dependence on Cooling Laser Parameters}
\label{subsec:intensity_noise}
To analyze the loss of data qubit coherence originating from \SI{435}{\nano\meter} light shift fluctuations, we observe the decoherence rate as a function of $\Omega_{435}$ and $\Delta_{435}$ as shown in Figure 7. For each setting of the \SI{435}{nm} laser power and detuning, the average decay rate of all the non-coolant ions is extracted respectively. From Figure 7a and 7b, the average decay rate obeys a quadratic dependence to $\Omega_{435}$ and an inverse linear dependence to $\Delta_{435}$. This qualitative relationship points to the existence of light shift, which is consistent with having a small, time-varying error on the laser amplitude stemming from intensity fluctuations.

We further investigate this by picking-off the beam before it addresses the ion and directing it on to a photodetector to measure its amplitude fluctuations. We observe the signal for a total duration of \SI{100}{\second} and measure the Allan deviation between consecutive segments for a range of time averaging periods between \SI{1}{\milli\second} to \SI{10}{\s}. This gives the relative amplitude fluctuation between segments for the total duration of the signal and covers all time-scales of interest in the experiment. Figure 8 shows the Allan deviation of the beam for a few different settings of the input optical power as well as the background signal without the \SI{435}{\nano\meter} beam. 
The Allan deviation for the beam follows the background signal initially and begins diverging at an averaging period of \SI{100}{\milli\second}, where relative amplitude drifts become important. Given the timescale of the qubit coherence measurements from Section \ref{subsec:coherence}, the Allan deviation between \SI{100}{\meter\second} to \SI{1}{\second} should be considered. The relative light shift induced from this fluctuation can be estimated by taking into account the normalized deviation measurement in this range. We assume a fractional time-varying intensity error of $\epsilon (t)$, resulting in a Rabi frequency of $\Omega(t) = \Omega_o\sqrt{1+\epsilon(t)}$. 
The relative phase accumulated between the two symmetric delay periods ($\tau/2$) within a shot can be expressed as $\Delta \phi = \frac{\Omega_0^2}{4\Delta}(\tau/2) - \frac{\Omega_0^2(1+\epsilon)}{4\Delta}(\tau/2) = \frac{\Omega_0^2}{8\Delta}\epsilon\tau$. The fidelity of the superposition state with this resultant phase error is $\lvert \langle + \vert e^{-i\Delta\phi\sigma_{z}/2} \vert + \rangle \rvert^2 = \cos^2 (\Delta\phi/2) = \frac{1}{2} (1+\cos(\Delta\phi))$. Taking the ensemble average over many shots yields $\langle \cos(\Delta\phi)\rangle = e^{-\frac{1}{2}{\langle \Delta \phi^2 \rangle}}$, in accordance with the ergodic theorem \cite{ergodic} and making the assumption that $\epsilon$ is a Gaussian random variable. Substituting the expression for the intra-shot relative phase from above gives a coherence decay with a functional form $e^{-\frac{\sigma^2_{\epsilon}}{2} \big(\frac{\Omega_0^2}{8\Delta}\tau \big)^2}$, where $\sigma_{\epsilon}$ is the Allan deviation associated with intensity error $\epsilon$. The coherence time can then be extracted, given by $\tau_c = \frac{8\sqrt{2}\Delta}{\Omega^2_o \sigma_{\epsilon}}$. We consider all possible transitions, including the side-bands, which contribute to the overall light shift. With the experimental parameters listed in Table 1, a coherence time of \SI{596}{\milli\second} to \SI{1145}{\milli\second} is obtained between a time averaging period of \SI{1}{\second} to \SI{100}{\milli\second} respectively, after adjusting for the COM mode duty cycle. This is approximately consistent with the coherence measurements in Figure 6.  

\subsection{Effect of Scattered 297 Photons}

To calculate the upper bound of the scattering rate, we assume that one sideband cooling pulse results in the emission of one 297 photon on the $ \ket{[3/2], 0} \rightarrow \ket{S, 1}$ transition. The number of
scattered photons absorbed by an ion spaced d = $\SI{5}{\mu\meter}$ in the chain per cooling pulse is as follows \cite{diff_isotope}.

\begin{equation}
    N_{297} = 6\pi \bigg(\frac{\lambda_{297}}{2\pi} \bigg)^2 \bigg(\frac{1}{4\pi d^2} \bigg) =1.34 \times 10^{-4}.
\end{equation}

Assuming 10 pulses/$\SI{10}{\milli\second}$, the maximum number of photons absorbed is 0.13 per second. 

The actual absorption rate in the experiment is even lower than the estimated value since this calculation relies on the fact that every cooling pulse causes a photon emission event. However, not each RSB pulse causes perfect population transfer to the $^2D_{3/2}$ state and one-third of the population still remains in one of the Zeeman level as explained in Section \ref{sec:setup}. Furthermore, we have also assumed an exact polarization match between the emitted photon and the ion absorbing it.

\begin{table}
    \centering
    \begin{minipage}{0.7\linewidth}
    \label{tab:exp_params}
    \begin{ruledtabular}
    \begin{tabular}{ll}
        \textbf{Parameter} & \textbf{Value} \\
        \hline
        Carrier Rabi frequency                                       & $2\pi \times \SI{125}{\kilo\hertz}$ \\
        Axial COM sideband Rabi frequency                            & $2\pi \times \SI{100}{\kilo\hertz}$ \\
        Other axial sideband Rabi frequency                          & $2\pi \times \SI{50}{\kilo\hertz}$  \\
        Radial sideband Rabi frequency                               & $2\pi \times \SI{10}{\kilo\hertz}$  \\
        Near-detuning                                                & $2\pi \times \SI{10}{\mega\hertz}$  \\
        Far-detuning                                                 & $2\pi \times \SI{30}{\mega\hertz}$  \\
        Allan deviation correction ($\tau=\SI{1}{\second}$)          & \SI{0.143}{\percent} \\
        Allan deviation correction ($\tau=\SI{100}{\milli\second}$)  & \SI{0.071}{\percent} \\
        Duty cycle (COM)                                             & \SI{70}{\percent} \\
    \end{tabular}
    \end{ruledtabular}
    \end{minipage}
    \caption{Experimental parameters used in the prediction of coherence times. Here, we include contributions from the off-resonant carrier, axial and radial side-band transitions. Near-detuning indicates the detuning between $\ket{S,1,\pm1} \leftrightarrow \ket{D,1, \mp 1}$ and the off-resonant transition stemming from the qubit state $\ket{1}$. Far-detuning takes into account the other Zeeman state, which is two Zeeman-shifts away from the former case.}
\end{table}

\end{document}